\documentclass[journal]{IEEEtran}

\ifCLASSINFOpdf
\else
   \usepackage[dvips]{graphicx}
\fi
\usepackage{url}

\usepackage{graphicx}
\usepackage{hyperref}
\usepackage{booktabs}
\usepackage{subcaption}
\usepackage{amsfonts}

\usepackage[backend=biber,style=numeric]{biblatex}
\begin{document}

\title{Diff-TONE: Timestep Optimization for iNstrument Editing in Text-to-Music Diffusion Models}

\author{Teysir Baoueb\textsuperscript{1}, Xiaoyu Bie\textsuperscript{1}, Xi Wang\textsuperscript{2}, Gaël Richard\textsuperscript{1}\\
\textsuperscript{1} LTCI, Télécom Paris, Institut Polytechnique de Paris, France\\
\textsuperscript{2} LIX, École Polytechnique, Institut Polytechnique de Paris, France

\thanks{This work was funded by the European Union (ERC, HI-Audio, 101052978). Views and opinions expressed are however those of the author(s) only and do not necessarily reflect those of the European Union or the European Research Council. Neither the European Union nor the granting authority can be held responsible for them.}
\thanks{This work was granted access to the HPC resources of IDRIS under the allocation 2024-AD011014567R1 made by GENCI.}
}
\maketitle

\begin{abstract}
Breakthroughs in text-to-music generation models are transforming the creative landscape, equipping musicians with innovative tools for composition and experimentation like never before. However, controlling the generation process to achieve a specific desired outcome remains a significant challenge. Even a minor change in the text prompt, combined with the same random seed, can drastically alter the generated piece. In this paper, we explore the application of existing text-to-music diffusion models for instrument editing. Specifically, for an existing audio track, we aim to leverage a pretrained text-to-music diffusion model to edit the instrument while preserving the underlying content. Based on the insight that the model first focuses on the overall structure or content of the audio, then adds instrument information, and finally refines the quality, we show that selecting a well-chosen intermediate timestep, identified through an instrument classifier, yields a balance between preserving the original piece's content and achieving the desired timbre. Our method does not require additional training of the text-to-music diffusion model, nor does it compromise the generation process's speed.
\end{abstract}

\IEEEpeerreviewmaketitle

\section{Introduction}\label{sec:introduction}
Recent advancements in deep learning and music generation models are providing musicians and composers with unprecedented tools for composition and experimentation~\cite{ma2024foundation}. In particular, recent development have led to significant progress in text-to-music generation~\cite{agostinelli2023musiclm,copet2023simple,lam2023efficient,liu2024audioldm,chen2024musicldm,melechovsky2024mustango,evans2024stableaudioopen} and timbre transfer models~\cite{cifka2021self,alinoori2022musicstar,comanducci2023timbre,li2024music,manor2024zero,novack2024ditto,rouard2024audio,tsai2024audio,zhang2024musicmagus}. These approaches enable the synthesis of expressive and diverse musical pieces from textual descriptions or the transformation of musical timbres while preserving content. These methods have already produced impressive results but suffer from a lack of flexibility and controllability. In fact, despite the recent progresses, controlling the generation process to achieve a specific desired outcome remains a major difficulty.  

Prior work in timbre transfer often necessitates training a separate model for each source-target timbre pair \cite{comanducci2023timbre, alinoori2022musicstar}. While effective for the predefined pairs, these approaches lack scalability when a large number of timbres need to be considered. To address this limitation, many-to-many timbre transfer methods have been explored. However, these methods \cite{comanducci2023timbre,baoueb2024wavetransfer} often rely on aligned datasets which are in practice difficult to obtain. Others, such as \cite{wu2023transplayer, cifka2021self, li2024music}, focus on learning disentangled representations for timbre and content, necessitating careful model design and optimization strategies.  
Parallel to these developments, text-to-music generation has seen remarkable growth, particularly through diffusion-based models. These models include text-to-music editing approaches that require specialized training such as \cite{tsai2024audio} with a specific learnable adapter or \cite{han2024instructme} which depends on paired source-target data for effective learning. 
A different direction was taken in MusicMagus~\cite{zhang2024musicmagus} where latent space manipulation for music editing is performed during inference but it relies on a music captioning model~\cite{liu2024music} and a large language model~\cite{chung2024scaling} to align text prompt distributions.

Concurrently, original strategies were proposed in \cite{rouard2024audio}  to condition a language model based music generation system with audio input. Impressive results are obtained but the models remain complex and a music language model needs to be trained from scratch for the most efficient approach.
Some other works aim at improving the controllability of the models by integrating content-based controls on innate music languages such as pitch or chords~\cite{lin2024content}. The model achieves flexible music variation generation and arrangement but, as underlined by the authors, does not fully meet semantic control expectations.

In this paper, we propose a novel strategy to improve the controllability and flexibility of text-to-music generation and timbre transfer. Given an existing audio track, our approach leverages a pretrained text-to-music diffusion model to modify the instrumentation while maintaining the structural and melodic integrity of the original piece. Our methodology is based on the observation that the diffusion model first establishes the overall structure and content of the audio, subsequently incorporates instrumental characteristics, and finally refines the quality. By identifying an optimal intermediate timestep through an instrument classifier, we demonstrate that our approach effectively balances content preservation with achieving the desired timbre transformation. Notably, our method does not require retraining the text-to-music diffusion model, nor does it introduce additional computational overhead to the generation process. Audio samples are provided at: \href{https://diff-tone.github.io/}{https://diff-tone.github.io}.

\section{Related Work}

\subsection{Diffusion Models and Text-to-Music Diffusion}
Diffusion models have emerged as the leading generative modeling paradigm, succeeding earlier influential frameworks such as Variational Autoencoders~\cite{kingma2014auto} and Generative Adversarial Networks~\cite{goodfellow2020generative}. Recently, diffusion models have demonstrated a remarkable ability to generate high-quality, diverse samples, achieving state-of-the-art performance across various generation tasks with minimal model collapse problem. Notable examples include image synthesis~\cite{song2021denoising,ho2020denoising}, text-to-image generation~\cite{dhariwal2021diffusion,rombach2022high}, video synthesis~\cite{blattmann2023stable, guo2024animatediff}, text-to-motion generation~\cite{tevet2023human,chen2023executing}, and audio generation~\cite{liu2023audioldm,huang2023make,ghosal2023text}. Similarly, text-to-music~\cite{agostinelli2023musiclm,huang2023noise2music}  models have emerged as a promising approach for generating music from textual descriptions by leveraging diffusion models, particularly latent diffusion. 

A notable example is Stable Audio Open Model \cite{evans2024stableaudioopen}, an open-weight text-to-audio generator by Stability AI. This model utilizes a latent diffusion framework to enable precise control over various musical attributes, including genre, tempo, and instrumentation. 

\subsection{Controllable and editable diffusion generation}
Making diffusion models controllable and aligned with human intentions requires conditioning the generation on user-specified inputs. Conditional diffusion models have been extensively studied~\cite{saharia2022photorealistic,ruiz2023dreambooth,balaji2022ediffi}. To enhance the fidelity of generated samples to given conditions and mitigate the common issue of condition-ignoring, two primary methods have been proposed: Classifier Guidance (CG)~\cite{dhariwal2021diffusion}, which employs an externally trained content classifier, and Classifier-Free Guidance (CFG)~\cite{ho2022classifier}, which leverages an implicit classifier through joint training of conditional and unconditional models with condition dropout. Recently, further studies have investigated the underlying mechanisms and behaviors of guidance methods. For instance, Kynk{\"a}{\"a}nniemi et al. \cite{kynkaanniemi2024applying} and Wang et al. \cite{wang2024analysis} demonstrate that certain timestep ranges are more informative for guiding the generation process, while Universal Guidance \cite{bansal2023universal} leverages off-the-shelf classifiers to guide generation specifically in image-related applications.

Also, in image processing, there are many examples underlining that, unlike generation tasks,  editing tasks focus on modifying specific aspects of existing images, such as regions or styles, utilizing generative models. Null-text inversion~\cite{mokady2023null} employs DDIM inversion~\cite{song2021denoising} and null-text optimization to facilitate text-based editing of real images. In contrast, InstructPix2Pix~\cite{brooks2023instructpix2pix} fine-tunes a diffusion model to follow user-provided editing instructions, incorporating both textual prompts and additional image guidance information.

\begin{figure*}[!ht]
    \centering
    \includegraphics[width=0.8\linewidth]{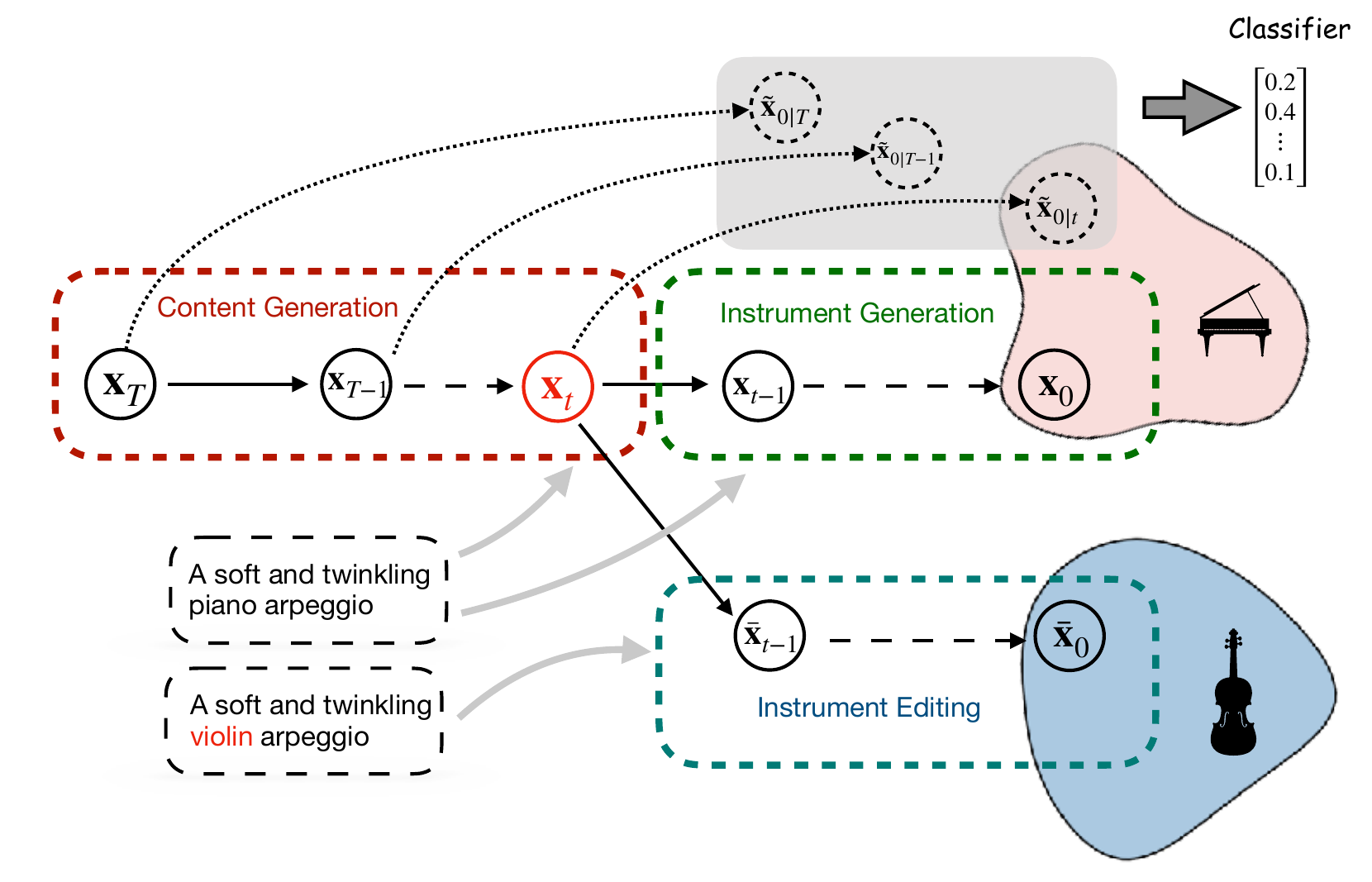}
    \caption{Pipeline of Diff-TONE: We start from an initial noisy latent $\mathbf{x}_T$ used to generate the audio with the original instrument. At each iteration of the inference process, we compute $\tilde{\mathbf{x}}_0$ and pass it to the classifier. Once the classifier does not change its prediction, we consider that to be the best time to substitute the target instrument for the initial instrument in the prompt. And we resume the generation process.}
    \label{fig:shared}
\end{figure*}

\section{Background}
\subsection{Mathematical Formulation of Diffusion Models}

Diffusion models transform random noise into content (e.g. music) that follows the data distribution. It usually involves a forward addition of noise and a backward denoising process, on either the data space (i.e., pixel for image)~\cite{ho2020denoising} or learnt latent spaces~\cite{rombach2022high}.

In the forward process, a data \(D_0\) is initially processed through a VAE-like encoder to obtain a latent embedding $\mathbf{x}_{0} = \mathcal{E}(D_{0})$, which is then progressively altered by adding random noise over $T$ steps:
\begin{equation}
\mathbf{x}_t = \sqrt{\alpha_t} \mathbf{x}_0 + \sqrt{1 - \alpha_t} \epsilon, \quad \epsilon \sim \mathcal{N}(0, I) \quad,
\label{eq1:forward_pass}
\end{equation}
where \(t \in [1, T]\) is the timestep, and \(\{\alpha_l\}_{l=1}^T\) is a scheduler that gradually decreases, controlling the extent of noise mixed with the data.

In the backward process, a network $\epsilon_\theta\left(\mathbf{x}_t, t, c\right)$ is trained to reverse this diffusion process by learning the added noise at each timestep \(t\), with $c$ as the condition, e.g., text embedding derived from the text prompt. Then the denoised embedding at timestep \(t-1\) can then be computed from previous timestep \(t\)s:
\begin{equation}
\mathbf{x}_{t-1} = \frac{\mathbf{x}_t - \left(\sqrt{1 - \alpha_t}\right) \epsilon_\theta\left(\mathbf{x}_t, t, c\right)}{\sqrt{\alpha_t}} \quad.
\label{eq3:denoise_process}
\end{equation}
The network parameters \(\theta\) are optimized by minimizing the Mean Squared Error (MSE) between the predicted noise and the ground truth~\cite{ho2020denoising}: $\theta^{'} = \arg\min_\theta \mathbb{E}_{\mathbf{x}_0, t, \epsilon \sim \mathcal{N}(0,1)}\left[\left\|\epsilon_\theta\left(\mathbf{x}_t, t, c\right) - \epsilon\right\|^2\right] \quad.$

\subsection{MERT Model}
The Music undERstanding model with large-scale self-supervised Training (MERT) \cite{li2024mert} is an acoustic music understanding model developed to address the challenges of modeling musical knowledge, particularly tonal and pitched characteristics. MERT incorporates teacher models to provide pseudo labels in a masked language modeling (MLM) style acoustic pre-training framework. Specifically, it utilizes an acoustic teacher based on Residual Vector Quantization - Variational AutoEncoder (RVQ-VAE) and a musical teacher based on the Constant-Q Transform (CQT). This combination has been found effective, outperforming conventional speech and audio approaches. The model scales from 95 million to 330 million parameters and has demonstrated strong generalization across 14 music understanding tasks, achieving state-of-the-art overall scores. The code and models are available online.

\section{Proposed Method}
In this work, we are interested in editing the timbre of music tracks generated by a text-to-music diffusion model.

During the inference process, we notice that text-to-music diffusion models focus on defining the overall structure of the audio, then adding timbre and finally refining the overall acoustic quality of the audio. Starting from this observation, we propose to tackle the task of timbre editing of an audio generated using a text-to-music model by swapping the instrument name in the text prompt at a meticulously chosen timestep. For this timestep, we should make sure that the content is well defined, but that timbre information has not been added yet. Given an instrument classifier, our approach is depicted in Figure~\ref{fig:shared} and consists of the following steps:
\setlength{\leftmargini}{14pt}
\begin{itemize}
    \item Set the random seed to the same one used to generate the audio with the initial instrument to get the same initial noisy latent $\mathbf{x}_T$.
    \item For each timestep $t: T \rightarrow 0$ of the generation process, we use an instrument classifier to detect whether instrument information has been embedded in the latent representation or not. Specifically, from the latent $\mathbf{x}_t$, we compute the predicted $\tilde{\mathbf{x}}_0$ defined as a linear prediction from the current step predicted noise $\theta(\mathbf{x}_t, t)$ towards the initial data, with $\bar{\alpha}_{t}=\prod_{l=1}^t \alpha_l$:
        \begin{equation}
            \tilde{\mathbf{x}}_0 = \frac{\mathbf{x}_t - \sqrt{1-\bar{\alpha}} \mathbf{\epsilon}_\theta(\mathbf{x}_t, t)}{\sqrt{\bar{\alpha}}}
        \end{equation}
    This quantity represents a rough but computationally-efficient approximation of $\mathbf{x}_0$ at each timestep, and thus is more suitable for instrument recognition than the noisy latent $\mathbf{x}_t$. The classifier might change its prediction during the generation process as it is less confident about its prediction at the beginning before becoming more and more confident as $t$ approaches the final inference timesteps. Thus, we consider the last time the classifier changes its prediction to be the timestep where instrument information starts being infused into the latent representation. For this timestep, we replace the instrument name in the text prompt with the target one, and continue the generation process.
\end{itemize}
We call our approach \textbf{Diff-TONE}, which stands for \textbf{T}imestep \textbf{O}ptimization for i\textbf{N}strument \textbf{E}diting in Text-to-Music \textbf{Diff}usion Models.

\subsection{Illustration of the Generation Process}
In this section, we show with an example how the diffusion process focuses on different aspects during the generation. For this purpose, we consider an audio having a single note generated with Stable Audio Open using the prompt: "A soft and elegant piano legato note.", and different timesteps $t \in \left[ 0, 49 \right] $ at which we swap the instrument name (piano) in the text prompt with the target one, which we choose to be violin here. Hence, the target prompt is: "A soft and elegant violin legato note.".

\begin{figure}[h!]
    \centering
    \begin{subfigure}{\linewidth}
        \centering
        \includegraphics[width=\linewidth]{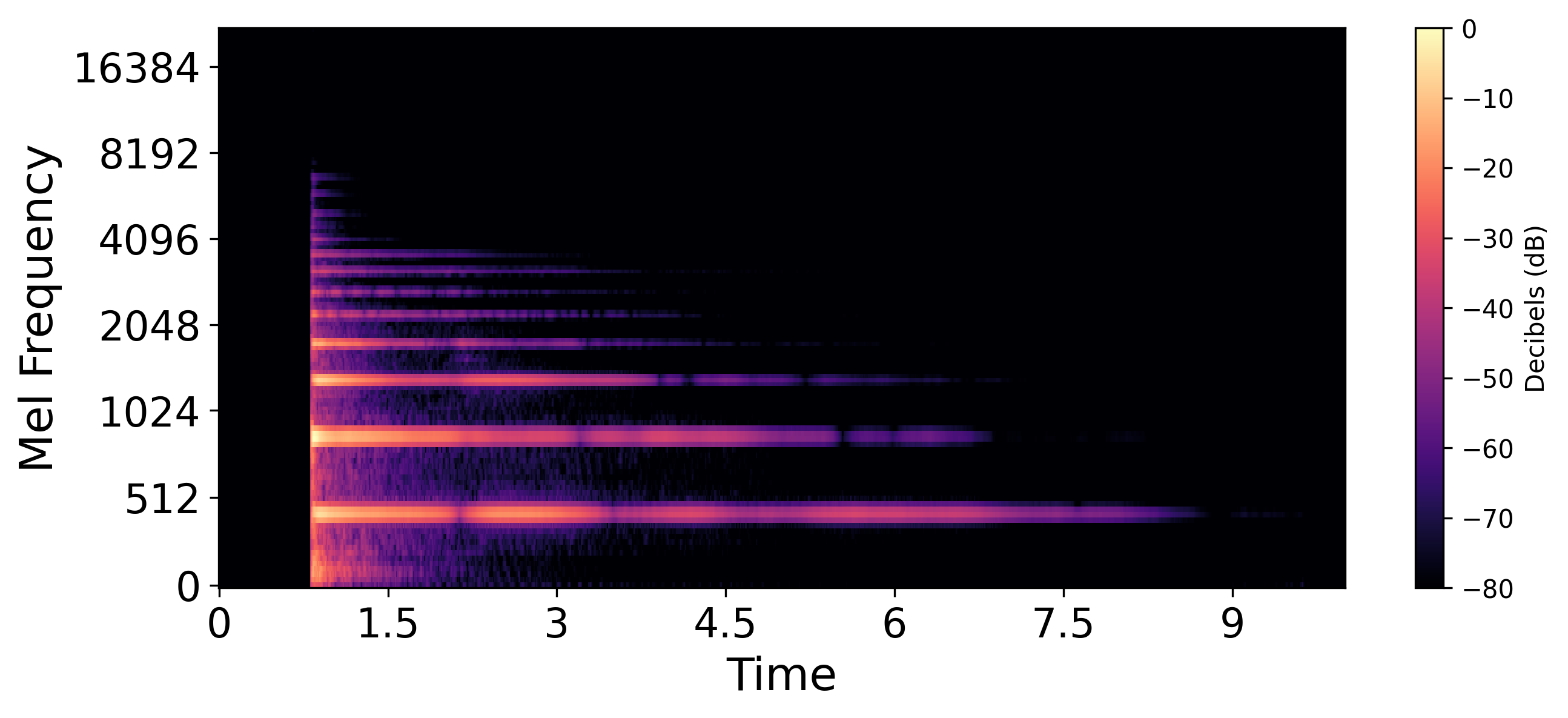}
        \caption{Original audio with no alterations to the inference process.}
        \label{fig:subfig1}
    \end{subfigure}
    
    \begin{subfigure}{\linewidth}
        \centering
        \includegraphics[width=\linewidth]{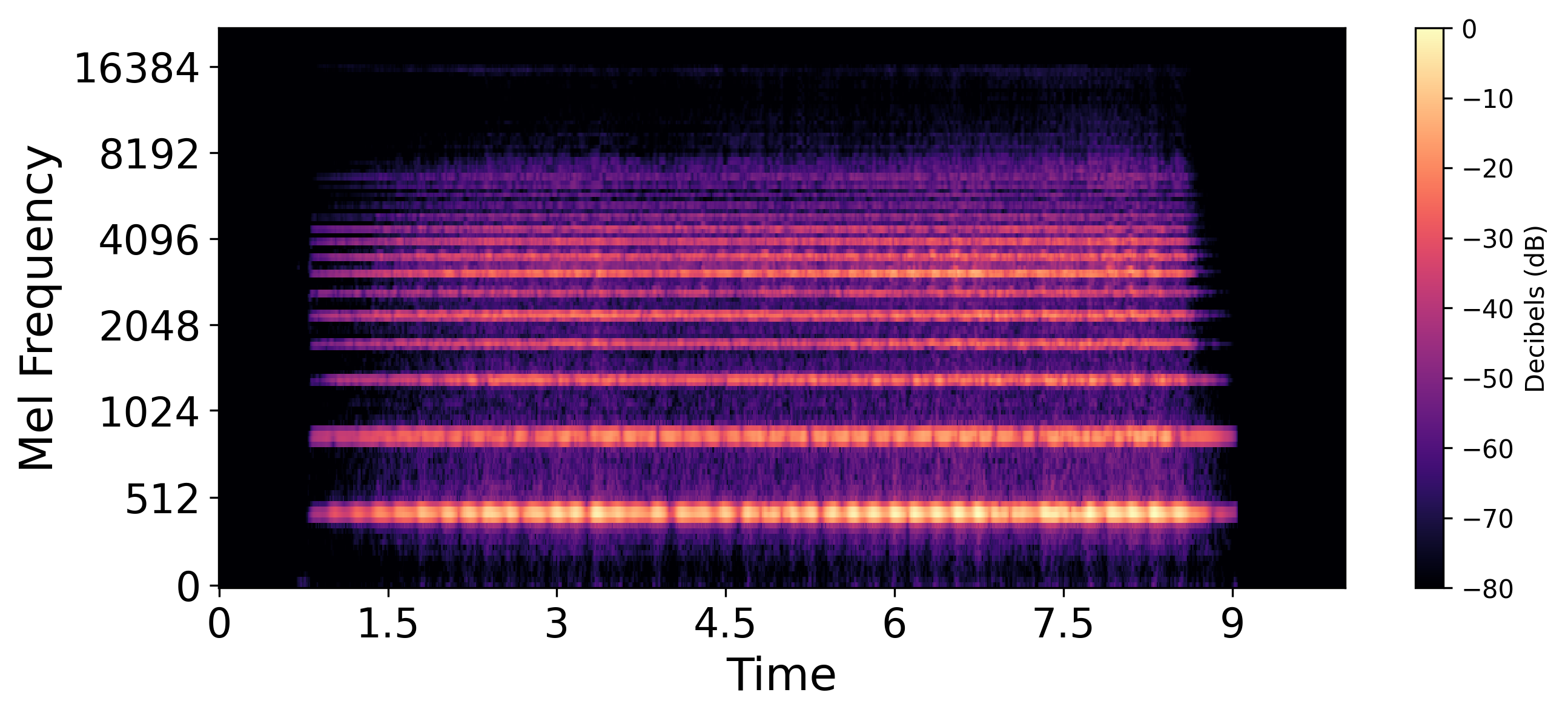}
        \caption{$t=39$}
        \label{fig:subfig2}
    \end{subfigure}
    
    \begin{subfigure}{\linewidth}
        \centering
        \includegraphics[width=\linewidth]{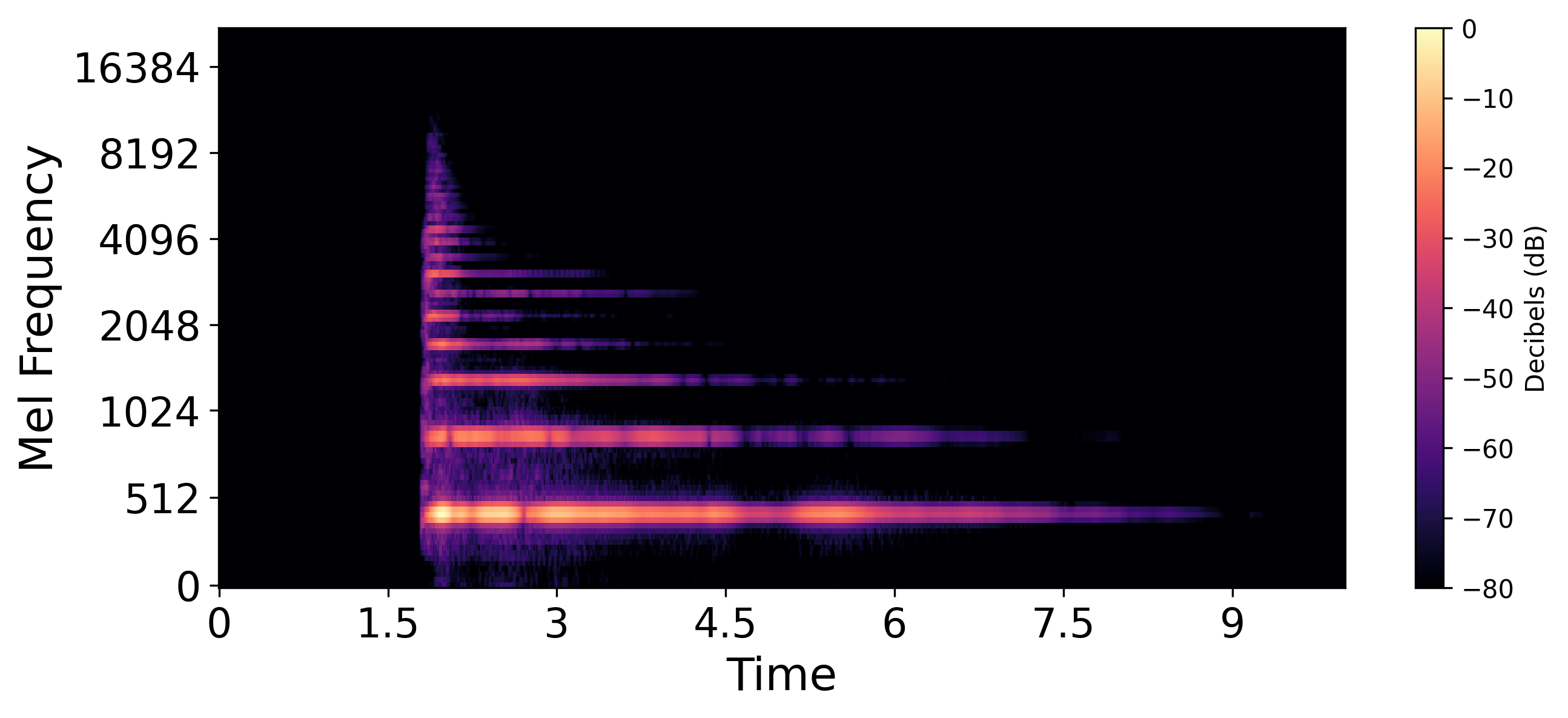}
        \caption{$t=21$}
        \label{fig:subfig3}
    \end{subfigure}
    
    \begin{subfigure}{\linewidth}
        \centering
        \includegraphics[width=0.95\linewidth]{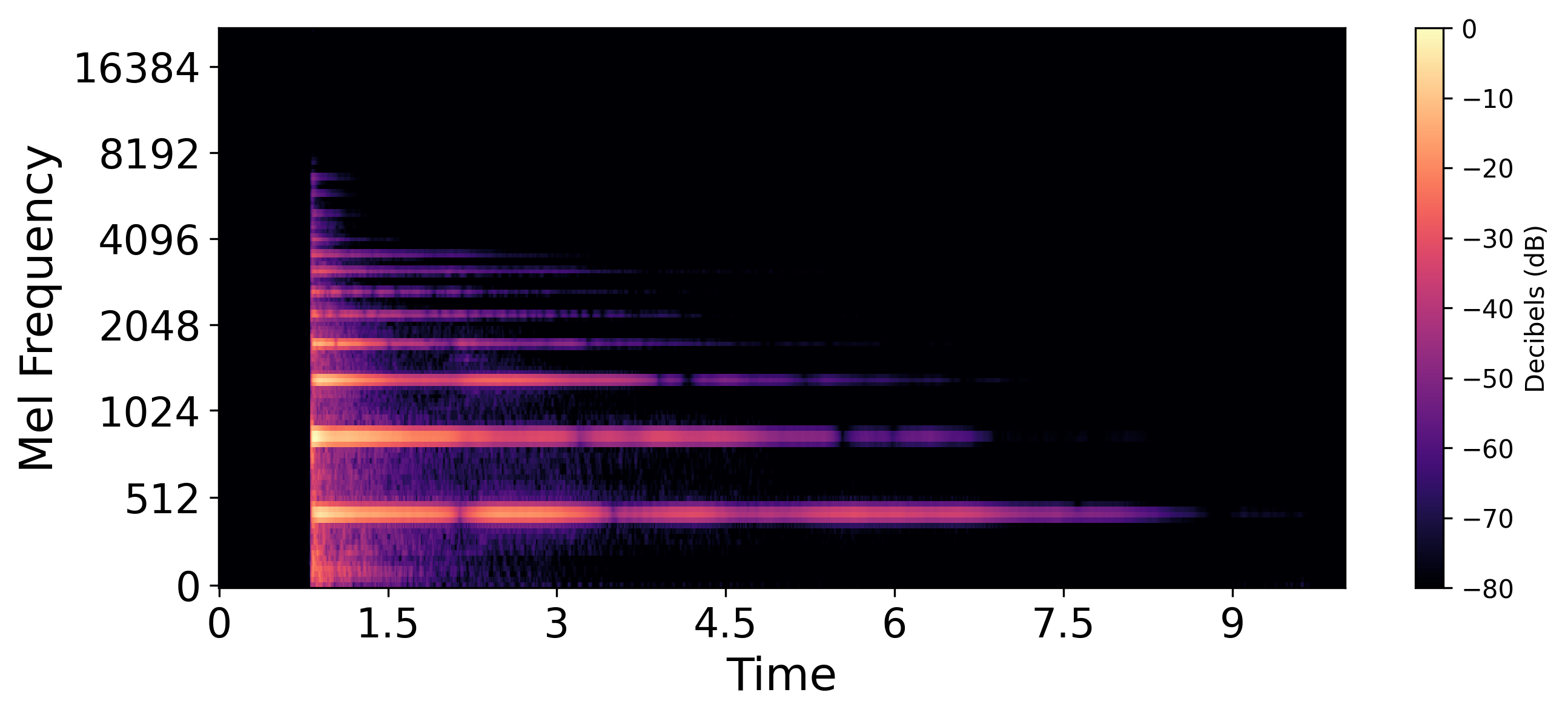}
        \caption{$t=9$}
        \label{fig:subfig4}
    \end{subfigure}

    \caption{Mel spectrogram corresponding to an audio generated with the prompt: "A soft and elegant piano legato note." and edited to "A soft and elegant violin legato note." at different timesteps.}
    \label{fig:mel_spectrogram}
\end{figure}

Figure \ref{fig:mel_spectrogram} shows the mel spectrogram for different settings. Figure \ref{fig:subfig1} corresponds to the audio generated without any alteration to the inference process. Figure \ref{fig:subfig2} showcases the result if we swap the instrument name at $t=39$. We notice that the note is different and has a vibrato effect. For a change at $t=21$ (Figure \ref{fig:subfig3}), the note is the same as the one played by the piano in the original audio, but with a violin timbre. In Figure \ref{fig:subfig4}, when we swap the prompts at $t=9$, nothing happens and we get the same audio as in Figure \ref{fig:subfig1}. This is one example that shows that the diffusion process focuses on the overall structure of the audio before infusing timbre information into it. In fact, when we try to change the instrument in the prompt at $t=9$, it fails as the timestep is too late and timbre is already embedded in the latent. The corresponding sound examples can be listened to on our demo page.

\subsection{Instrument Classifier}
To classify instruments, one option is to use an off-the-shelf 
classifier. However, this would significantly increase computational load, as we would need to decode the predicted $\tilde{\mathbf{x}}_0$ representation at each diffusion timestep before classification. To mitigate this, we train a classifier directly on the latent representation obtained from the Stable Audio Open encoder.

To obtain a rich and meaningful representation for instrument classification, we distill the MERT model \cite{li2024mert}. Since the MERT representation (13 layers, time dimension, feature dimension = 768) is significantly larger than that of the Stable Audio Open encoder (feature dimension = 64, time dimension), we first downsample the teacher model's representation to a more compact yet informative form.

To achieve this, we follow a three-stage training process. In the first stage, we train a classifier using cross-entropy loss on the MERT representation. This classifier includes an aggregation layer to select the most relevant representation layer(s), a linear layer to reduce the time dimension to a fixed size of $16$, another linear layer to reduce the feature dimension to $512$, and a classification head composed of linear layers, a dropout layer, and an LSTM layer.

After completing this stage, we discard the classification head and use its input as the teacher representation for distillation. The student network architecture consists of an average pooling layer that reduces the time dimension to $128$, followed by the Mimi architecture \cite{defossez2024moshi} with three blocks. In each block, the number of feature channels is doubled while the time dimension is halved. To ensure the student network effectively captures the teacher's representation, we use a loss function based on cosine similarity. Finally, once the student network is trained and frozen, we train a classification head with the same architecture as before on top of it.

\section{Experiments}
\subsection{IRMAS Dataset}
IRMAS (Instrument Recognition in Musical Audio Signals) dataset \cite{bosch2012comparison} is a collection of $11000$ audio recordings provided in WAV format, with a sampling rate of $44.1$ kHz. It contains 11 instrument classes (cello, clarinet, flute, acoustic guitar, electric guitar, organ, piano, saxophone, trumpet, violin, and human singing voice) recorded in various real-world musical contexts. Since we are interested in instrumental music, we exclude the human singing voice class from IRMAS.

\subsection{Evaluation Set}\label{sec:eval_set}
We generate $10$ prompts for each of the $10$ classes in the IRMAS dataset using ChatGPT, resulting in a total of $100$ prompts. These prompts are used both to create an evaluation set for classification and to perform timbre editing, where the duration of each generated audio is $10$ seconds.

\subsection{Instrument Classifier Training Details}
For distillation and classification, we use the IRMAS dataset, splitting it into $93.2$\% for training and $7.8$\% for validation. We first train a classifier on top of the MERT representation for $50$ epochs. Then, we perform distillation for another $50$ epochs, followed by training a classifier on the student representation for an additional $50$ epochs.

All training is done using the Adam optimizer and takes approximately $1.5$ hours on a single A$40$ GPU.

To evaluate the effectiveness of the distillation process, we also train a classifier directly on the Stable Audio Open representation (without distillation) using the same architecture and training it for $50$ epochs.
\subsection{Baselines}
We evaluate our approach against several baseline methods. The first, Random Timestep Selection (Diff-Random), involves swapping the text prompt at a randomly chosen timestep during the generation process. The second, Middle Timestep (Diff-Midpoint), performs the prompt change at the midpoint of the timesteps. Finally, we compare our method to MusicMagus \cite{zhang2024musicmagus}, a recent zero-shot text-to-music editing model.

\subsection{Metrics}
To evaluate the different methods, we employ the following objective metrics:

\begin{itemize}
\item \textbf{Chromagram Similarity (Chroma):} A chromagram represents the tonal content of an audio signal. By comparing the chromagrams of the timbre-edited audio and the original audio, we assess how well the tonal characteristics are preserved.
\item \textbf{Top-1 Instrument Accuracy (Inst. Acc.):} To determine whether the generated audio matches the target instrument, we use the classifier we trained on top of the teacher model (MERT representation) on IRMAS dataset. This metric measures whether the predicted instrument class aligns with the intended target. 
\item \textbf{Kernel Audio Distance (KAD):} KAD \cite{chung2025kad} is a computationally efficient, distribution-free metric that is based on Maximum Mean Discrepancy (MMD) and used for audio quality evaluation. We use CLAP-LAION music embeddings and the IRMAS Training set as the reference set for each instrument.
\end{itemize}

\section{Results}
\subsection{Classification Results}
The first-stage classification in the MERT model achieves an accuracy of 72\%, which is comparable to the results reported in \cite{li2024mert} for instrument classification on other datasets and aligns with state-of-the-art performance on the IRMAS dataset \cite{tiemeiger2024towards}.

The classification accuracy on the evaluation set generated using Stable Audio Open is presented in Table \ref{tab:classification_results}.  Notably, the classifier trained directly on the latent representation from Stable Audio Open performs poorly on out-of-domain data. In contrast, the classifier fine-tuned through the distillation process yields significantly better results, demonstrating improved generalization.

It is important to highlight that the generated evaluation set includes some waveforms that do not accurately reflect the instrument specified in the text prompt, e.g., a piano instead of an organ. This discrepancy is one of the reasons we opted against training on audios generated by Stable Audio Open, as manual curation is required to ensure correct instrument labeling.

\begin{table}[h!]
\centering
    \begin{tabular}{c|c}
    \toprule
    \textbf{Classifier} & \textbf{Top-1 Accuracy}\\
    \midrule
    Teacher Classifier & $\mathbf{53\%}$\\
    Distilled Classifier & $38\%$\\
    Non-distilled Classifier & $10\%$\\
    \bottomrule
    \end{tabular}
\caption{Classification accuracy on a set of 100 audio samples generated with Stable Audio Open. \textbf{Teacher classifier}: trained on top of the MERT representation given raw waveform as input (upper bound). \textbf{Distilled classifier}: trained on top of the distilled model which is trained by aligning the output of the student network with MERT features given latent representation as input. \textbf{Non-distilled classifier}: trained directly on the latent representation of the Stable Audio Open encoder without distillation.}
\label{tab:classification_results}
\end{table}

\subsection{Timbre Editing Results}
\subsubsection{Comparison with other methods for selecting the timestep for instrument swap}
In this section, we consider timbre editing results between $90$ instrument pairs\footnote{We have $10$ instruments in the dataset and we have $9$ different target instruments for each.}. For each pair, we consider $10$ prompts generated by ChatGPT for the initial instrument as mentioned in Section \ref{sec:eval_set}. Table~\ref{tab:timbre_results} presents the results comparing our Diff-TONE method with two alternative prompt-change timing strategies: Diff-Random, which selects a random timestep for applying the prompt change, and Diff-Midpoint, which applies the change at the midpoint timestep. We can see that our proposed approach yields better results on two objective metrics: Chroma and KAD. Specifically, it exhibits a reduced chromagram distance, demonstrating superior content preservation. The KAD shows better overall generation quality. For the top-1 instrument accuracy, we explain the fact that the midpoint approach has better results by the fact that the midpoint timestep would lead to a timbre change in the audio more often than not. At this stage, the instrument information has typically not yet been fully embedded in the latent representation. As a result, modifying the prompt at this point is more likely to successfully alter the instrument in the generated audio. However, this improvement comes at the cost of reduced content fidelity, as the audio content may undergo significant changes, as indicated by the chromagram similarity metric.

As mentioned above, the instrument classifier used was trained on the IRMAS dataset. Nevertheless, we observed similar trends in the classification results (although with lower top-1 accuracy for all methods) with the checkpoint of a classifier trained on MTG-Jamendo on top of the MERT representation (provided by the authors of MERT).

\begin{table}[h!]
\centering
\resizebox{\linewidth}{!}{
    \begin{tabular}{c|c|c|c}
    \toprule
    \textbf{Method} & \textbf{Chroma} ($\downarrow$) & \textbf{KAD} ($\downarrow$) & \textbf{Inst. Acc.} ($\uparrow$) \\
    \midrule
    Diff-Random & $0.148$ & $18.846$ & $28.89\%$\\
    Diff-Midpoint & $0.189$ & $20.716$ & $\mathbf{39.33}\%$\\
    Diff-TONE & $\mathbf{0.099}$ & $\mathbf{18.265}$ & $23.00\%$ \\
    \bottomrule
    \end{tabular}
}
\caption{Timbre editing results for the different approaches used to select a timestep for instrument swap in the prompt}
\label{tab:timbre_results}
\end{table}

\subsubsection{Comparison to MusicMagus}

We also compare our approach more specifically to MusicMagus on the same three specific word swapping pairs used in their original paper, namely piano $\rightarrow$ acoustic guitar, piano $\rightarrow$ organ, and violin $\rightarrow$ piano. Our model obtains better results on all aspects which underlines the merits of our strategy for music timbre editing.

\begin{table}[h!]
\centering
\resizebox{\linewidth}{!}{
    \begin{tabular}{c|c|c|c}
    \toprule
    \textbf{Method} & \textbf{Chroma} ($\downarrow$) & \textbf{KAD} ($\downarrow$) & \textbf{Inst. Acc.} ($\uparrow$) \\
    \midrule
    MusicMagus \cite{zhang2024musicmagus}& $0.216$ & $33.154$ & $23.33\%$\\
    Diff-TONE & $\mathbf{0.108}$ & $\mathbf{30.815}$ & $\mathbf{26.66\%}$ \\
    \bottomrule
    \end{tabular}
}
\caption{Timbre editing results on these three pairs: piano $\rightarrow$ acoustic guitar, piano $\rightarrow$ organ, and violin $\rightarrow$ piano}
\label{tab:timbre_results_2}
\end{table}

In addition to these objective evaluations, a large number of sound examples are provided in our demo page for these specific pairs but also for a variety of other pairs to give a better view of the merits and limits of our method.

For reproducibility purposes, our complete code will be made available upon acceptance. 

\section{Conclusion}
In this paper, we explore text-to-music diffusion models for instrument editing. Given an input audio track, we leverage a pretrained model to modify the instrument while preserving the original content. By identifying an optimal intermediate timestep via an instrument classifier, our method balances content preservation and timbre control. Notably, it requires no additional training and maintains generation efficiency. 
Though, our approach faces some limitations. For instance, the classifier may never change its prediction from the start of the generation process which prevents us from finding a suitable  timestep for editing. Also, in some other cases the timestep selected by the classifier is too late into the generation process, which implies that almost no editing is performed. 
Future work will be dedicated to the improvement of the robustness of our approach. Other perspectives of our work include its application to real audio data and its extension to the simultaneous editing of multiple instruments.

\printbibliography

@string{icassp="Proc. ICASSP"}

@string{mlsp="Proc. MLSP"}

@string{ismir="Proc. ISMIR"}

@string{icml="Proc. ICML"}

@string{iclr="Proc. ICLR"}

@string{neurips="Proc. NeurIPS"}

@string{ijcai="Proc. IJCAI"}

@string{jmlr="JMLR"}

@string{cvpr="Proc. CVPR"}

@string{tmlr = "Trans. Mach. Learn. Res."}

@string{aaai = "Proc. AAAI"}

@string{ijcai = "Proc. IJCAI"}

@string{TASLP_acm = "IEEE/ACM Trans. Audio, Speech, Language Process."}

@string{NAACL = "Proc. NAACL"}

@article{ma2024foundation,
  title={Foundation models for music: A survey},
  author={Ma, Yinghao and {\O}land, Anders and Ragni, Anton and Del Sette, Bleiz MacSen and Saitis, Charalampos and Donahue, Chris and Lin, Chenghua and Plachouras, Christos and Benetos, Emmanouil and Shatri, Elona and others},
  journal={arXiv preprint arXiv:2408.14340},
  year={2024}
}

@article{agostinelli2023musiclm,
  title={Musiclm: Generating music from text},
  author={Agostinelli, Andrea and Denk, Timo I and Borsos, Zal{\'a}n and Engel, Jesse and Verzetti, Mauro and Caillon, Antoine and Huang, Qingqing and Jansen, Aren and Roberts, Adam and Tagliasacchi, Marco and others},
  journal={arXiv preprint arXiv:2301.11325},
  year={2023}
}

@inproceedings{copet2023simple,
  title={Simple and controllable music generation},
  author={Copet, Jade and Kreuk, Felix and Gat, Itai and Remez, Tal and Kant, David and Synnaeve, Gabriel and Adi, Yossi and D{\'e}fossez, Alexandre},
  booktitle=neurips,
  year={2023}
}

@article{lam2023efficient,
  title={Efficient neural music generation},
  author={Lam, Max WY and Tian, Qiao and Li, Tang and Yin, Zongyu and Feng, Siyuan and Tu, Ming and Ji, Yuliang and Xia, Rui and Ma, Mingbo and Song, Xuchen and others},
  journal=neurips,
  year={2023}
}

@article{liu2024audioldm,
  title={Audioldm 2: Learning holistic audio generation with self-supervised pretraining},
  author={Liu, Haohe and Yuan, Yi and Liu, Xubo and Mei, Xinhao and Kong, Qiuqiang and Tian, Qiao and Wang, Yuping and Wang, Wenwu and Wang, Yuxuan and Plumbley, Mark D},
  journal=TASLP_acm,
  year={2024},
  publisher={IEEE}
}

@inproceedings{chen2024musicldm,
  title={Musicldm: Enhancing novelty in text-to-music generation using beat-synchronous mixup strategies},
  author={Chen, Ke and Wu, Yusong and Liu, Haohe and Nezhurina, Marianna and Berg-Kirkpatrick, Taylor and Dubnov, Shlomo},
  booktitle=ICASSP,
  year={2024},
}

@inproceedings{melechovsky2024mustango,
  title={Mustango: Toward controllable text-to-music generation},
  author={Melechovsky, Jan and Guo, Zixun and Ghosal, Deepanway and Majumder, Navonil and Herremans, Dorien and Poria, Soujanya},
  booktitle=NAACL,
  year={2024}
}

@inproceedings{manor2024zero,
  title={Zero-shot unsupervised and text-based audio editing using DDPM inversion},
  author={Manor, Hila and Michaeli, Tomer},
  booktitle=icml,
  year={2024}
}

@inproceedings{novack2024ditto,
  title={Ditto: Diffusion inference-time t-optimization for music generation},
  author={Novack, Zachary and McAuley, Julian and Berg-Kirkpatrick, Taylor and Bryan, Nicholas J},
  booktitle=icml,
  year={2024}
}

@inproceedings{li2024mert,
  title={{MERT}: Acoustic Music Understanding Model with Large-Scale Self-supervised Training},
  author={Yizhi LI and Ruibin Yuan and Ge Zhang and Yinghao Ma and Xingran Chen and Hanzhi Yin and Chenghao Xiao and Chenghua Lin and Anton Ragni and Emmanouil Benetos and Norbert Gyenge and Roger Dannenberg and Ruibo Liu and Wenhu Chen and Gus Xia and Yemin Shi and Wenhao Huang and Zili Wang and Yike Guo and Jie Fu},
  booktitle=ICLR,
  year={2024},
}

@inproceedings{evans2024stableaudioopen,
  title={Stable audio open},
  author={Evans, Zach and Parker, Julian D and Carr, CJ and Zukowski, Zack and Taylor, Josiah and Pons, Jordi},
  booktitle=ICASSP,
  year={2025}
}

@inproceedings{bosch2012comparison,
  title={A Comparison of Sound Segregation Techniques for Predominant Instrument Recognition in Musical Audio Signals.},
  author={Bosch, Juan J and Janer, Jordi and Fuhrmann, Ferdinand and Herrera, Perfecto},
  booktitle=ismir,
  year={2012}
}

@article{defossez2024moshi,
  title={Moshi: a speech-text foundation model for real-time dialogue},
  author={D{\'e}fossez, Alexandre and Mazar{\'e}, Laurent and Orsini, Manu and Royer, Am{\'e}lie and P{\'e}rez, Patrick and J{\'e}gou, Herv{\'e} and Grave, Edouard and Zeghidour, Neil},
  journal={arXiv preprint arXiv:2410.00037},
  year={2024}

}

@article{chung2025kad,
  title={KAD: No More FAD! An Effective and Efficient Evaluation Metric for Audio Generation},
  author={Chung, Yoonjin and Eu, Pilsun and Lee, Junwon and Choi, Keunwoo and Nam, Juhan and Chon, Ben Sangbae},
  journal={arXiv preprint arXiv:2502.15602},
  year={2025}
}

@inproceedings{zhang2024musicmagus,
  title={Musicmagus: Zero-shot text-to-music editing via diffusion models},
  author={Zhang, Yixiao and Ikemiya, Yukara and Xia, Gus and Murata, Naoki and Mart{\'\i}nez-Ram{\'\i}rez, Marco A and Liao, Wei-Hsiang and Mitsufuji, Yuki and Dixon, Simon},
  booktitle=IJCAI,
  year={2024}
}

@inproceedings{alinoori2022musicstar,
  title={Music-STAR: a Style Translation system for Audio-based Re-instrumentation}, 
  author={Alinoori, Mahshid and Tzerpos, Vassilios},
  year={2022},
  booktitle=ismir
}

@inproceedings{comanducci2023timbre,
  title={Timbre transfer using image-to-image denoising diffusion implicit models}, 
  author={Luca Comanducci and Fabio Antonacci and Augusto Sarti},
  year={2023},
  booktitle=ismir
}

@inproceedings{baoueb2024wavetransfer,
  title={WaveTransfer: A Flexible End-to-end Multi-instrument Timbre Transfer with Diffusion},
  author={Baoueb, Teysir and Bie, Xiaoyu and Janati, Hicham and Richard, Gaël},
  booktitle=MLSP,
  year={2024}
}

@inproceedings{wu2023transplayer,
  author={Wu, Yuxuan and He, Yifan and Liu, Xinlu and Wang, Yi and Dannenberg, Roger B.},
  booktitle=icassp, 
  title={Transplayer: Timbre Style Transfer with Flexible Timbre Control}, 
  year={2023},
}

@INPROCEEDINGS{cifka2021self,
  author={Cífka, Ondřej and Ozerov, Alexey and Şimşekli, Umut and Richard, Gaël},
  booktitle=icassp, 
  title={Self-Supervised VQ-VAE for One-Shot Music Style Transfer}, 
  year={2021},
}

@inproceedings{tsai2024audio,
  title={Audio Prompt Adapter: Unleashing Music Editing Abilities for Text-to-Music with Lightweight Finetuning}, 
  author={Fang-Duo Tsai and Shih-Lun Wu and Haven Kim and Bo-Yu Chen and Hao-Chung Cheng and Yi-Hsuan Yang},
  year={2024},
  booktitle=ismir
}

@inproceedings{han2024instructme,
  title={InstructME: An Instruction Guided Music Edit And Remix Framework with Latent Diffusion Models}, 
  author={Bing Han and Junyu Dai and Weituo Hao and Xinyan He and Dong Guo and Jitong Chen and Yuxuan Wang and Yanmin Qian and Xuchen Song},
  year={2024},
  booktitle=ijcai,
}

@inproceedings{li2024music,
  title={Music style transfer with time-varying inversion of diffusion models},
  author={Li, Sifei and Zhang, Yuxin and Tang, Fan and Ma, Chongyang and Dong, Weiming and Xu, Changsheng},
  booktitle=AAAI,
  year={2024}
}

@inproceedings{rouard2024audio,
  title={Audio Conditioning for Music Generation via Discrete Bottleneck Features}, 
  author={Simon Rouard and Yossi Adi and Jade Copet and Axel Roebel and Alexandre Défossez},
  year={2024},
  booktitle=ismir
}

@inproceedings{lin2024content,
  title={Content-based controls for music large language modeling},
  author={Lin, Liwei and Xia, Gus and Jiang, Junyan and Zhang, Yixiao},
  booktitle=ismir,
  year={2024}
}

@article{goodfellow2020generative,
  title={Generative adversarial networks},
  author={Goodfellow, Ian and Pouget-Abadie, Jean and Mirza, Mehdi and Xu, Bing and Warde-Farley, David and Ozair, Sherjil and Courville, Aaron and Bengio, Yoshua},
  journal={Commun. ACM},
  volume={63},
  number={11},
  pages={139--144},
  year={2020}
}

@inproceedings{kingma2014auto,
  title={Auto-encoding variational bayes},
  author={Kingma, Diederik P and Welling, Max and others},
  year={2014},
  booktitle=iclr
}

@article{blattmann2023stable,
  title={Stable video diffusion: Scaling latent video diffusion models to large datasets},
  author={Blattmann, Andreas and Dockhorn, Tim and Kulal, Sumith and Mendelevitch, Daniel and Kilian, Maciej and Lorenz, Dominik and Levi, Yam and English, Zion and Voleti, Vikram and Letts, Adam and others},
  journal={arXiv preprint arXiv:2311.15127},
  year={2023}
}

@inproceedings{song2021denoising,
  title={Denoising Diffusion Implicit Models},
  author={Song, Jiaming and Meng, Chenlin and Ermon, Stefano},
  booktitle=iclr,
  year={2021}
}

@inproceedings{ho2020denoising,
  title={Denoising diffusion probabilistic models},
  author={Ho, Jonathan and Jain, Ajay and Abbeel, Pieter},
  booktitle=neurips,
  year={2020}
}

@inproceedings{rombach2022high,
  title={High-resolution image synthesis with latent diffusion models},
  author={Rombach, Robin and Blattmann, Andreas and Lorenz, Dominik and Esser, Patrick and Ommer, Bj{\"o}rn},
  booktitle=cvpr,
  year={2022}
}

@inproceedings{dhariwal2021diffusion,
  title={Diffusion models beat gans on image synthesis},
  author={Dhariwal, Prafulla and Nichol, Alexander},
  booktitle=neurips,
  year={2021}
}

@inproceedings{guo2024animatediff,
  title={Animatediff: Animate your personalized text-to-image diffusion models without specific tuning},
  author={Guo, Yuwei and Yang, Ceyuan and Rao, Anyi and Liang, Zhengyang and Wang, Yaohui and Qiao, Yu and Agrawala, Maneesh and Lin, Dahua and Dai, Bo},
  booktitle=iclr,
  year={2024}
}

@inproceedings{tevet2023human,
  title={Human motion diffusion model},
  author={Tevet, Guy and Raab, Sigal and Gordon, Brian and Shafir, Yonatan and Cohen-Or, Daniel and Bermano, Amit H},
  booktitle=iclr,
  year={2023}
}

@inproceedings{chen2023executing,
  title={Executing your Commands via Motion Diffusion in Latent Space},
  author={Chen, Xin and Jiang, Biao and Liu, Wen and Huang, Zilong and Fu, Bin and Chen, Tao and Yu, Gang},
  booktitle=cvpr,
  year={2023},
}

@inproceedings{saharia2022photorealistic,
  title={Photorealistic text-to-image diffusion models with deep language understanding},
  author={Saharia, Chitwan and Chan, William and Saxena, Saurabh and Li, Lala and Whang, Jay and Denton, Emily L and Ghasemipour, Kamyar and Gontijo Lopes, Raphael and Karagol Ayan, Burcu and Salimans, Tim and others},
  booktitle=neurips,
  year={2022}
}

@inproceedings{ruiz2023dreambooth,
  title={Dreambooth: Fine tuning text-to-image diffusion models for subject-driven generation},
  author={Ruiz, Nataniel and Li, Yuanzhen and Jampani, Varun and Pritch, Yael and Rubinstein, Michael and Aberman, Kfir},
  booktitle=cvpr,
  year={2023}
}

@article{balaji2022ediffi,
  title={ediffi: Text-to-image diffusion models with an ensemble of expert denoisers},
  author={Balaji, Yogesh and Nah, Seungjun and Huang, Xun and Vahdat, Arash and Song, Jiaming and Kreis, Karsten and Aittala, Miika and Aila, Timo and Laine, Samuli and Catanzaro, Bryan and others},
  journal={arXiv preprint arXiv:2211.01324},
  year={2022}
}

@inproceedings{ho2022classifier,
  title={Classifier-Free Diffusion Guidance},
  author={Ho, Jonathan and Salimans, Tim},
  booktitle={Proc. NeurIPS Workshops},
  year={2021}
}

@inproceedings{kynkaanniemi2024applying,
  title={Applying guidance in a limited interval improves sample and distribution quality in diffusion models},
  author={Kynk{\"a}{\"a}nniemi, Tuomas and Aittala, Miika and Karras, Tero and Laine, Samuli and Aila, Timo and Lehtinen, Jaakko},
  booktitle=neurips,
  year={2024}
}

@article{wang2024analysis,
  title={Analysis of Classifier-Free Guidance Weight Schedulers},
  author={Wang, Xi and Dufour, Nicolas and Andreou, Nefeli and Cani, Marie-Paule and Abrevaya, Victoria Fern{\'a}ndez and Picard, David and Kalogeiton, Vicky},
  journal=tmlr,
  year={2024}
}

@inproceedings{bansal2023universal,
  title={Universal guidance for diffusion models},
  author={Bansal, Arpit and Chu, Hong-Min and Schwarzschild, Avi and Sengupta, Soumyadip and Goldblum, Micah and Geiping, Jonas and Goldstein, Tom},
  booktitle=cvpr,
  year={2023}
}

@inproceedings{mokady2023null,
  title={Null-text inversion for editing real images using guided diffusion models},
  author={Mokady, Ron and Hertz, Amir and Aberman, Kfir and Pritch, Yael and Cohen-Or, Daniel},
  booktitle=cvpr,
  year={2023}
}

@inproceedings{brooks2023instructpix2pix,
  title={Instructpix2pix: Learning to follow image editing instructions},
  author={Brooks, Tim and Holynski, Aleksander and Efros, Alexei A},
  booktitle=cvpr,
  year={2023}
}

@article{chung2024scaling,
  title={Scaling instruction-finetuned language models},
  author={Chung, Hyung Won and Hou, Le and Longpre, Shayne and Zoph, Barret and Tay, Yi and Fedus, William and Li, Yunxuan and Wang, Xuezhi and Dehghani, Mostafa and Brahma, Siddhartha and others},
  journal=jmlr,
  volume={25},
  number={70},
  pages={1--53},
  year={2024}
}

@inproceedings{liu2024music,
  title={Music understanding llama: Advancing text-to-music generation with question answering and captioning},
  author={Liu, Shansong and Hussain, Atin Sakkeer and Sun, Chenshuo and Shan, Ying},
  booktitle=ICASSP,
  year={2024}
}

@INPROCEEDINGS{tiemeiger2024towards,
  author={Tiemeijer, Paul and Shahsavari, Mahyar and Fazlali, Mahmood},
  booktitle={2024 IEEE International Conference on Omni-layer Intelligent Systems (COINS)}, 
  title={Towards Music Instrument Classification using Convolutional Neural Networks}, 
  year={2024},
  volume={},
  number={},
  pages={1-6},
  doi={10.1109/COINS61597.2024.10622136}}

@article{liu2023audioldm,
  title={Audioldm: Text-to-audio generation with latent diffusion models},
  author={Liu, Haohe and Chen, Zehua and Yuan, Yi and Mei, Xinhao and Liu, Xubo and Mandic, Danilo and Wang, Wenwu and Plumbley, Mark D},
  journal={arXiv preprint arXiv:2301.12503},
  year={2023}
}

@inproceedings{huang2023make,
  title={Make-an-audio: Text-to-audio generation with prompt-enhanced diffusion models},
  author={Huang, Rongjie and Huang, Jiawei and Yang, Dongchao and Ren, Yi and Liu, Luping and Li, Mingze and Ye, Zhenhui and Liu, Jinglin and Yin, Xiang and Zhao, Zhou},
  booktitle={International Conference on Machine Learning},
  pages={13916--13932},
  year={2023},
  organization={PMLR}
}

@inproceedings{ghosal2023text,
  title={Text-to-audio generation using instruction guided latent diffusion model},
  author={Ghosal, Deepanway and Majumder, Navonil and Mehrish, Ambuj and Poria, Soujanya},
  booktitle={Proceedings of the 31st ACM International Conference on Multimedia},
  pages={3590--3598},
  year={2023}
}

@article{huang2023noise2music,
  title={Noise2music: Text-conditioned music generation with diffusion models},
  author={Huang, Qingqing and Park, Daniel S and Wang, Tao and Denk, Timo I and Ly, Andy and Chen, Nanxin and Zhang, Zhengdong and Zhang, Zhishuai and Yu, Jiahui and Frank, Christian and others},
  journal={arXiv preprint arXiv:2302.03917},
  year={2023}
}

\end{document}